\documentclass[UTF8]{article}
\usepackage{style_files/ijcai22}

\usepackage{times}

\usepackage{soul}
\usepackage{url}
\usepackage[hidelinks]{hyperref}
\usepackage[utf8]{inputenc}
\usepackage[small]{caption}
\usepackage{graphicx}
\usepackage{pdfpages}
\usepackage{amsmath}
\usepackage{booktabs}
\usepackage{refcount}
\usepackage{amsfonts}
\usepackage{subfig}
\usepackage{caption}
\usepackage{color, colortbl}
\usepackage{wrapfig}
\usepackage{breqn}
\usepackage{enumerate}
\usepackage{color}
\usepackage{threeparttable}
\usepackage{tabularx}

\usepackage[noend]{algpseudocode}
\usepackage{algorithmicx,algorithm}

\usepackage{amsmath,amssymb,amsfonts}
\usepackage{algpseudocode}
\usepackage{indentfirst} 
\usepackage{graphicx}
\usepackage{float}
\usepackage{framed}
\usepackage{caption}
\usepackage{textcomp}
\usepackage{xcolor}
\usepackage{tipa}
\usepackage{tabularx}
\usepackage{gensymb}
\usepackage{multirow}
\usepackage{makecell}
\usepackage{array}
\usepackage{booktabs}
\usepackage{threeparttable}
\usepackage{pifont}
\usepackage{titlesec}
\usepackage{enumitem}
\usepackage{appendix}
\usepackage{stfloats}

\newtheorem{problem}{Problem}

\newtheorem{pro-stat}{Problem Definition}

\def\T{{\scriptscriptstyle\mathsf{T}}}

\newcommand{\hide}[1]{}

\newcommand{\suger}{\textsc{SuGeR}}

\title{\suger: A Subgraph-based Graph Convolutional Network Method for Bundle Recommendation}

\author{
Zhenning Zhang$^1$\and
Boxin Du$^1$\and
Hanghang Tong$^1$\footnote{Contact Author}\\
\affiliations
$^1$University of Illinois at Urbana-Champaign
\emails
\{zz45,boxindu2 ,htong\}@illinois.edu
}

\begin{document}
\setlength{\abovedisplayskip}{1.4pt}
\setlength{\belowdisplayskip}{1.4pt}
\setlength{\abovedisplayshortskip}{1.4pt}
\setlength{\belowdisplayshortskip}{1.4pt}
\newcommand\mycommfont[1]{\small\ttfamily\textcolor{blue}{#1}}
\maketitle


\begin{abstract}
Bundle recommendation is an emerging research direction in the recommender system with the focus on recommending customized bundles of items for users. Although Graph Neural Networks (GNNs) have been applied in this problem and achieve superior performance, existing methods underexplore the graph-level GNN methods, which exhibit great potential in traditional recommender system. Furthermore, they usually lack the transferability from one domain with sufficient supervision to another domain which might suffer from the label scarcity issue. In this work, we propose a subgraph-based Graph Neural Network model, \suger, for bundle recommendation to handle these limitations. \suger\ generates heterogeneous subgraphs around the user-bundle pairs, and then maps those subgraphs to the users' preference predictions via neural relational graph propagation. Experimental results show that \suger\ significantly outperforms the state-of-the-art baselines in both the basic and the transfer bundle recommendation problems. 
\end{abstract}

\section{Introduction}

\noindent Bundle recommendation is a newly emerging research direction in the recommender system. Generally, bundle recommendation aims to recommend a bundle of items that collectively might be more appealing to users, compared with recommending single items. For example, large online game and music distributors such as Steam, Tencent, Netease, and large e-commerce platforms such as Amazon and Taobao have already begun to sell their products in bundles \cite{bai2019personalized,deng2020personalized,liang2019rec} . This recommendation and selling strategy would benefit both sellers and their customers mutually. 

However, despite its importance, less effort has been devoted to this direction than the traditional user-item recommendation problem. Existing approaches for traditional recommender systems tend to fall short for the bundle recommendation problem for two main reasons. First, the bundle recommendation problem contains three types of interactions: user-item preference, user-bundle preference, and bundle-item membership. It is non-trivial for traditional user-item recommendation methods to incorporate various types of interactions. Second, the label scarcity problem and the cold-start problem for certain domains are crucial as a newly rising direction. One possible solution is to learn and transfer knowledge from other domains with sufficient supervision. However, most traditional methods do not have transferability between various domains. 

Recently, there are sparse literatures on studying the bundle recommendation problem, and they demonstrate the potential strengths of adopting Graph Neural Networks (GNNs) on this problem \cite{chang2021rec,xian2021ex3,chang2021bundle} . However, these methods bear some fundamental limitations as follows. First, almost all of the current GNN-based methods adopt node-level GNNs to learn the embeddings for users and bundles. But the graph-level GNN is underexplored, which can automatically learn the heuristic rules for prediction, and also shows significant performance improvements for traditional recommender systems \cite{zhang2019inductive}. Leveraging the graph-level GNN in the bundle recommendation scenario is still an open question. Second, the label leakage issue for bundle recommendation is often neglected. But it has a significant impact in terms of overfitting and performance. Third, like the traditional recommender system, most current methods could not be effectively generalized to unseen data or even transferred from one domain to another. 

In this paper, we propose a \underline{Su}bgraph-based \underline{G}raph Convolutional N\underline{e}two\underline{r}k model (\suger), to tackle all of the aforementioned limitations. The key high-level idea is to construct a heterogeneous subgraph for each user-bundle pair and map it to predict the user's preference via graph-level GNNs. Our method can handle both the basic bundle recommendation problem and the transfer bundle recommendation problem, in which the goal is to apply the model learned on one domain to another. In summary, the contributions of the paper are as follows. 

\begin{itemize}
    \item \textbf{Novel Problem Setting.} We define two problem settings in the paper and develop a model to tackle both. We are the first to study the transfer bundle recommendation problem to our best knowledge. 
    \item \textbf{Subgraph-based GNN Model.} We propose \suger, a subgraph-based Graph Convolutional Network for two bundle recommendation problem settings. 
    \item \textbf{Empirical Evaluations.} We conduct extensive empirical evaluations to demonstrate the effectiveness of \suger, which significantly outperforms the current state-of-the-art by all metrics in both problem settings. 
\end{itemize}

\section{Problem Definition and Preliminaries}\label{sec:problem}

\noindent In this section, we start with the formal definition of the bundle recommendation problem, followed by the preliminaries of related recent works on GNN-based matrix completion.

\subsection{Problem Definition}

\noindent We use $\mathcal{U}, \mathcal{I}, \mathcal{B}$ to denote user set, item set and bundle set for training and test. Then, we use $|\mathcal{U}|, |\mathcal{I}|, |\mathcal{B}|$ to denote the number of bundles, items and users. We define $\mathbf{X}, \mathbf{Y}, \mathbf{Z}$ as user-bundle, user-item, and bundle-item interaction matrices respectively (the left side of Figure \ref{tab:overall_graph}). $\mathbf{X}$ and $\mathbf{Y}$ represent users' bundle and item preference, and $\mathbf{Z}$ represents bundle-item membership. Taking $\mathbf{X}$ as an example. $\mathbf{X}(u,b) = 1$ if there exists observed interactions between user $u$ and bundle $b$ where $u\in \mathcal{U}, b\in \mathcal{B}$. $\mathbf{Y}$ and $\mathbf{Z}$ follow similar definitions. 
These three matrices then be split into training and test sets. Based on these fundamental symbol definitions, the basic bundle recommendation problem is defined as follows:


\begin{problem}{\textbf{Basic Bundle Recommendation:}}

\textbf{Given:} The user set $\mathcal{U}$, item set $\mathcal{I}$ and bundle set $\mathcal{B}$ with their corresponding observed user-bundle, user-item, and bundle-item interaction matrices $\mathbf{X}, \mathbf{Y}, \mathbf{Z}$ for training.
    
\textbf{Output:} The predicted user-bundle preference score matrix $\mathbf{R}\in\mathbb{R}^{|\mathcal{U}|\times|\mathcal{B}|}$ for all unobserved user-bundle pairs. $\mathbf{R}(u, b)$ is a real number in $(0,1)$ indicating the probability of user $u$ favors bundle $b$. 
\end{problem}

Furthermore, we also study the problem of transfer bundle recommendation, where the goal is to learn a user-bundle encoder on the source domain and then apply the learned encoder on the target domain. The transfer bundle recommendation problem is formally defined as follows. 
\begin{problem}{\textbf{Transfer Bundle Recommendation:}}

\textbf{Given:} The user set $\mathcal{U}_s$, item set $\mathcal{I}_s$ and bundle set $\mathcal{B}_s$ with their corresponding observed user-bundle, user-item, and bundle-item interaction matrices $\mathbf{X}_s, \mathbf{Y}_s, \mathbf{Z}_s$ for training from the source domain. The user set $\mathcal{U}_t$, item set $\mathcal{I}_t$ and bundle set $\mathcal{B}_t$ with their corresponding observed user-bundle, user-item, and bundle-item interaction matrices $\mathbf{X}_t, \mathbf{Y}_t, \mathbf{Z}_t$ for test from the target domain. 

\textbf{Output:} The predicted user-bundle preference score matrix $\mathbf{R}\in\mathbb{R}^{|\mathcal{U}_t|\times|\mathcal{B}_t|}$ for all of the unobserved user-bundle pairs in the target domain $\{\mathcal{U}_t, \mathcal{I}_t, \mathcal{B}_t\}$. $\mathbf{R}(u, b)\in(0,1)$ indicating the probability of user $u\in\mathcal{U}_t$ favors bundle $b\in\mathcal{B}_t$.

\end{problem}
Note that the observed user-bundle pairs in the target domain are divided into training and test splits where there is no overlap. We only use training split as the observed pairs, and the test split is used for test. 


\subsection{Preliminaries}
\noindent In this subsection we briefly introduce a GNN-based inductive matrix completion model, namely IGMC \cite{zhang2019inductive} since our \suger\ model is inspired by IGMC. 
\par IGMC chooses RGCN \cite{schlichtkrull2017modeling} as the basic layer. The basic idea is two-fold. First, for a given rating, the model extracts an enclosing subgraph around the user-item pair and then uses the extracted subgraph for information propagation. Second, the extracted enclosing subgraph is mapped to a prediction score of the centered user-item pair based on a neural model, including a GNN layer. 
\par The key of IGMC is to adopt the graph-level GNN and map the extracted enclosing subgraphs to the user-item scores, which can automatically learn suitable heuristics for the user-item recommendation. However, IGMC is not designed for the more complicated bundle recommendation scenario. Furthermore, it is still unknown how to adopt the graph-level GNN for the bundle recommendation problem. The first distinctive difference of the proposed \suger\ is that it generates $k$-hop heterogeneous subgraph from the inputs of user-bundle, user-item, and bundle-item interaction as model input. Secondly, in the bundle context, we consider the weight of bundles in the information propagation process, adding similarity factor and try to split RGCN into two-level propagation. We also consider the label leakage issue and propose a solution. Therefore, compared to IGMC, \suger\ is a bundle-oriented recommendation model.

\section{The Proposed Model}\label{sec:method}

\begin{figure*}[htbp]
    \small
    \centering
    \includegraphics[width=18cm]{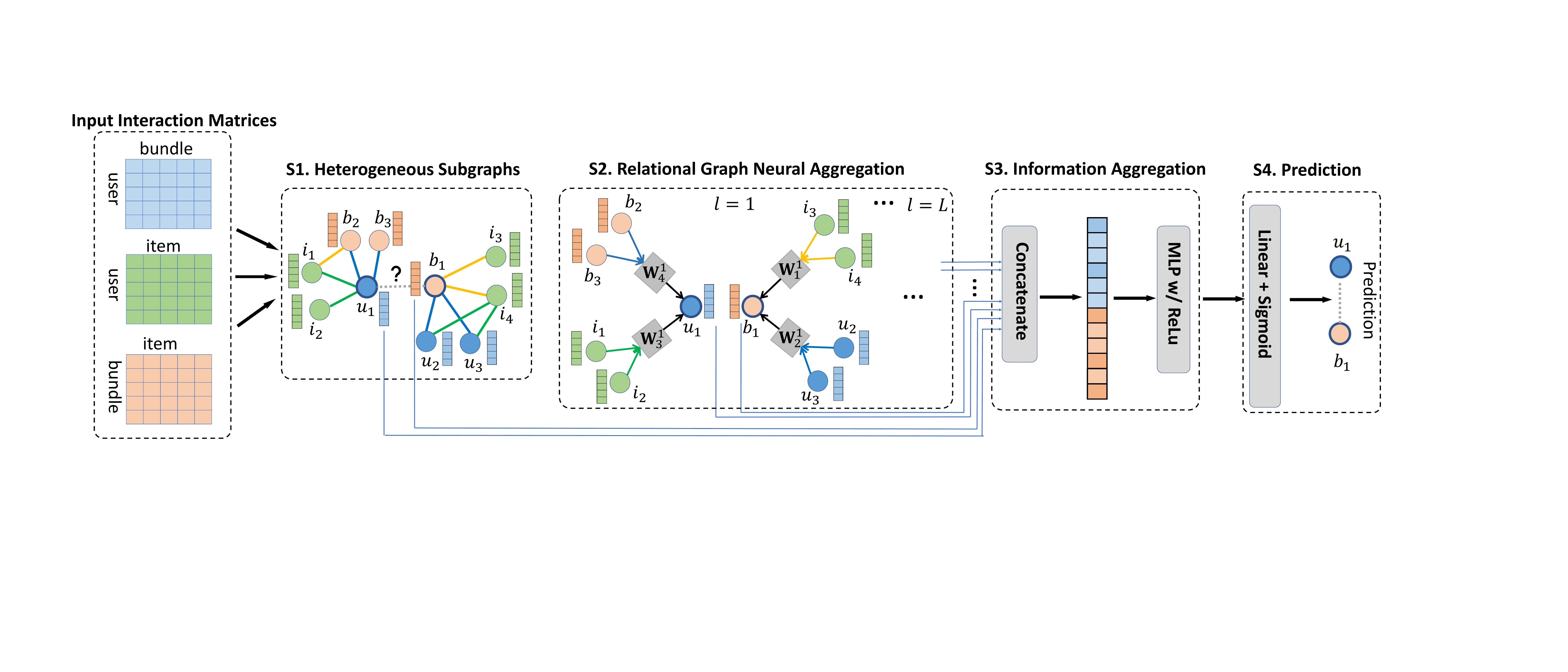} 
    \vspace{-3.0\baselineskip}
    \caption{The overall pipeline of \suger\ model. The whole procedure could be separated into 4 stages. Here we use the centered user-bundle pair ($u_1, b_1$) as an example, and we only show 1-hop
    heterogeneous subgraph. The concrete illustration is presented in Section \ref{sec:method}.}
       \label{tab:overall_graph}
\end{figure*}

\noindent The overall model pipeline is shown in Figure \ref{tab:overall_graph}. There are four main stages which are S1. Heterogeneous subgraph generation and feature initialization, S2. Relational graph neural propagation, S3. Information aggregation, and S4. Prediction. The main steps are summarized as follows. First, in S1, we construct the $k$-hop heterogeneous subgraphs for given user-bundle pairs from the training user-bundle, user-item interaction matrices, and training bundle-item membership matrix. Then we initialize the embeddings for the bundle, item, and user from category and feature information. Second, we adopt relational GNN layers on the extracted heterogeneous subgraphs, which can be seen as a two-level neural propagation process. The two levels of information propagation by \suger\ model will capture the internal structure of the user's preference on items and bundles and also the influence of items on bundles. Third, we concatenate initial and intermediate embeddings of all relational GNN layers as information aggregation method and use MLP with Sigmoid function as prediction output. This section also illustrates the label leakage issue and its impact on the bundle recommendation problem.


\subsection{Heterogeneous subgraph generation and feature initialization}\label{sec:initialization}

\noindent To adopt the graph-level GNN model, there are several methods for graph construction, such as hypergraph, homogeneous graph, and heterogeneous graph. Here, we use heterogeneous graph for the subgraph construction for user-bundle pairs. Compared to homogeneous graph, heterogeneous graph could model different levels of information propagations between different categories of nodes and edges.
\par The heterogeneous subgraph we construct is an directed graph $\mathcal{G}_s$ = \{$\mathcal{V}$, $\mathcal{E}$\}. For each user-bundle pair, we generate one $k$-hop heterogeneous subgraph. We divide nodes $ \textit{v} \in  \mathcal{V} $ into several types. For the centered user-bundle pair in a subgraph, we mark centered user as 0 and centered bundle as 1. Then, we assign several types for the non-centered user, bundle and item. For bundle, user and item in hop $k_{i}$, we assign categorical attributes $3k_{i} - 1, 3k_{i}, 3k_{i} + 1$ to them. The reason of categorizing nodes in this way is that for different hops the nodes have different influence on the centered u-b pair, so we assign different types for users, items and bundles in each hop $k$. For instance, when $k$ = 1, there are five types of nodes in total. 
\par For edges, we model edges $ \textit{e}\in\mathcal{E} $ into six categories: user$\rightarrow$bundle and bundle$\rightarrow$user (if $\mathbf{X}(u,b) = 1$) , user$\rightarrow$item and item$\rightarrow$user (if $\mathbf{Y}(u,i) = 1$), bundle$\rightarrow$item and item$\rightarrow$bundle (if $\mathbf{Z}(b,i) = 1$). In the directed heterogeneous graph, edges related to user have different semantics from bundle-item affiliation edges and we hope the parameters for user propagation are different from bundle propagation.
\par Based on the above directed graph, we denote matrices of bundle embeddings, item embeddings and user embeddings respectively as $\mathbf{S}\in\mathbb{R}^{|\mathcal{B}|\times d}, \mathbf{D}\in\mathbb{R}^{|\mathcal{I}|\times d}, \mathbf{F}\in\mathbb{R}^{|\mathcal{U}|\times d}$. Here $d$ is the dimension of embeddings. Note that different from previous bundle recommendation works, the user embeddings may differ when we consider different user-bundle pairs since each heterogeneous subgraph could have a different embedding for the centered user and the bundle embeddings. 

\begin{center}
$ \mathbf{e}_{b} = \mathbf{S}^\T \mathbf{v}_{b}, \mathbf{e}_{i} = \mathbf{D}^\T \mathbf{v}_{i}, \mathbf{e}_{u} = \mathbf{F}^\T \mathbf{v}_{u}, $
\end{center}

\par where $\mathbf{v}_{b}\in\mathbb{R}^{|\mathcal{B}|\times 1}, \mathbf{v}_{i}\in\mathbb{R}^{|\mathcal{I}|\times 1}, \mathbf{v}_{u} \in\mathbb{R}^{|\mathcal{U}|\times 1} $, representing the one-hot encoding for bundles, items and users. In the training process, the embedding matrices $\mathbf{S, D, F}$ will be updated iteratively. 

\par For the embedding initialization, we fuse type and feature to generate initial embeddings. We use one-hot encoding for node types to obtain the first vector. We randomly generate Gaussian distribution feature vector for another half of the initial embedding. The reason for using Gaussian distribution here is similar to the idea of free embeddings from \cite{wu2020diffnet++}, which could serve as a regularization for the embedding learning. Finally, we concatenate these two vectors to form an initial embedding for every node. 
\par One crucial characteristic of \suger\ is its transfer learning ability. There are two main reasons which are closely related to the subgraph generation and feature initialization. First, by using heterogeneous subgraph as input, \suger\ tries to learn a subgraph encoder to map the user-bundle centered subgraph into embedding space. Some similar graph structures might be shared by the data from different domains, which our subgraph encoder could capture. Such graph structures are closely correlated with users' buying habits which is independent of dataset domains. Furthermore, our subgraph encoder focuses on the local graph pattern near a user or a bundle, while existing methods (e.g., BGCN \cite{chang2021bundle}) focus on learning a global embedding of users/bundles when predicting a particular user-bundle preference. For the transfer learning problem, it might not be reasonable for encoders to learn the whole graph pattern dependent on the dataset domain. Second, no side information such as users' ages and genders is used during initialization, because the model intends to learn domain-independent features. 

\subsection{Relational graph neural propagation}\label{sec:propagation}
\noindent In this section, we introduce the structure of our \suger\ model and elaborate on the details of information propagation. The basic layer component in our GNN model is RGCN layer  \cite{schlichtkrull2017modeling}, which is developed specifically to deal with the highly multi-relational data characteristic of realistic knowledge bases and heterogeneous graph operations. The propagation model is as follows:
\begin{equation}\label{eq:rgcn}
\mathbf{e}_{i}^{(l+1)}=\sigma\left(\sum_{r \in \mathcal{P}} \sum_{j \in \mathcal{N}_{i}^{r}} \frac{1}{c_{i, r}} \mathbf{W}_{r}^{(l)} \mathbf{e}_{j}^{(l)}+\mathbf{W}_{0}^{(l)} \mathbf{e}_{i}^{(l)}\right)
\end{equation}

\noindent where $\mathcal{P}$ is the node type set and $\mathcal{N}_{i}^{r}$ is the $r$ type neighbor of node $i$, and $\mathbf{W}_{r}^{(l)}$ is the weight matrix for type $r$ at layer $l$. $c_{i, r}$ is a problem-specific regularization factor,  we use the number of neighbors $\left|\mathcal{N}_{r}(i)\right|$ in \suger\ context to regularize the influence by each type since we do not want the number of nodes strongly affects the propagation process. 


\par We illustrate the information propagation path and why it is suitable for the bundle recommendation from two perspectives.

\par Firstly, let us discuss information propagation from the bundle's perspective. Intuitively, items in a bundle have certain connections with each other. For example, it is less likely for a game shop to combine a game DVD with a toilet seat to form a bundle. Furthermore, the customers who buy bundles could influence bundles' semantics. \suger\ provides a weight matrix to capture this internal relationship between user and bundle. As a result, for bundles, there are two levels of propagation ($i\rightarrow b$, $u \rightarrow b$) as follows:

\begin{small}
\begin{equation}\nonumber
\begin{aligned}
&\mathbf{e}_{b,1}^{(\ell+1)}=\frac{1}{\left|\mathcal{N}_{i,b}\right|} \sigma\left(\mathbf{W}_{1}^{(\ell)}\left(\mathbf{e}_{b}^{(\ell)}+\tau\left(\{\mathbf{e}_{i}^{(\ell)} \mid i \in \mathcal{N}_{i,b}\}\right)\right)
 +\mathbf{c}_{1}^{(\ell)}\right) \\
 &\mathbf{e}_{b,2}^{(\ell+1)}=\frac{1}{\left|\mathcal{N}_{u,b}\right|} \sigma\left(\mathbf{W}_{2}^{(\ell)}\left(\mathbf{e}_{u}^{(\ell)}+\tau\left(\{\mathbf{e}_{u}^{(\ell)} \mid u \in \mathcal{N}_{u,b}\}\right)\right)
 +\mathbf{c}_{2}^{(\ell)}\right) 
\end{aligned}
\end{equation}
\end{small}

Here, $\mathbf{e}_{b,1}^{(\ell+1)}$ and $\mathbf{e}_{b,2}^{(\ell+1)}$ are the two parts of bundle embeddings in the $\left(\ell+1\right)$th layer, and $W_{1}^{(\ell)}, W_{2}^{(\ell)}, $ are the two updating weight matrices based on edge type in heterogeneous graph. The reason why there are two parts is we want to gather information from user and item nodes to refine bundle embeddings. $\tau(\cdot)$ is the aggregation function, here we simply choose sum and $\sigma(\cdot)$ is \textit{LeakyRelu} activation function. After activation, we regularize them based on neighbor number. 

\par Secondly, we consider the user's perspective. The bundle can be regarded as several organized items to be sold to the user, so the user's preference towards one bundle will be influenced by items in bundle. For centered and non-centered bundle nodes, the information will be gathered and aggregated from the connected user and item. A bundle is comprised of several items (they have their information propagation channels), and user's decision on a bundle will not only be affected by item itself but also by the power when items appear as a set. For instance, one user may not like chocolate and tea individually, but when they appear together, user will buy them since tea can counteract the sweetness of chocolate.
Another critical factor is that by comparing the containing items of two bundles, we can roughly get the similarity between them. The propagation layer will generate similar embedding in the characteristic space of two very identical bundles to recommend similar bundles to a user. To let \suger\ learn the semantic of item combination and similarity information between bundles, we have the following two equations:

\begin{small}
\begin{equation}\nonumber
\begin{aligned}
&\mathbf{e}_{u,1}^{(\ell+1)}=\frac{1}{\left|\mathcal{N}_{i,u}\right|} \sigma\left(\mathbf{W}_{3}^{(\ell)}\left(\mathbf{e}_{u}^{(\ell)}+\tau\left(\{\mathbf{e}_{i}^{(\ell)} \mid i \in \mathcal{N}_{i,u}\}\right)\right)
 +\mathbf{c}_{3}^{(\ell)}\right) \\
&\mathbf{e}_{u, 2}^{(\ell+1)}=\frac{1}{\left|\mathcal{N}_{b,u}\right|}\sigma\left(\mathbf{W}_{4}^{(\ell)}\left(\mathbf{e}_{u}^{(\ell)}+\tau\left(\{\delta_{b}\mathbf{e}_{b}^{(\ell)} \mid b \in \mathcal{N}_{b,u}\}\right)\right)+\mathbf{c}_{4}^{(\ell)}\right)
\end{aligned}
\end{equation}
\end{small}

\par Here $\mathbf{e}_{u,1}^{(\ell+1)}$ and $\mathbf{e}_{u,2}^{(\ell+1)}$ are two parts of user embedding in the $\left(\ell+1\right)$th layer, and $\mathbf{W}_{3}^{(\ell)}, \mathbf{W}_{4}^{(\ell)}, $ are the two updating weight matrices for item$\rightarrow$user and bundle$\rightarrow$user edges. $\tau(\cdot)$ is the aggregation function, and here we still choose sum. $\sigma(\cdot)$ is \textit{LeakyRelu} activation function. One thing different from bundle's perspective is we apply factor $\delta_{b} = 1 + \eta_{b}$ to involve the extra weights between bundles based on similarity during propagation. 
Among all bundles there is one bundle ${e}_{t}$ that best reflects user's preference. We try to find ${e}_{t}$ and give it more weight. Before the $\left(\ell+1\right)$th iteration, we concatenate $\mathbf{e}_{b}^{(\ell)}$ to itself for five times to get $\mathbf{e}_{bf}^{(\ell)}$ for every bundle $b$ in subgraph. We do the same to $\mathbf{e}_{u}^{(\ell)}$ to get $\mathbf{e}_{uf}^{(\ell)}$. $\mathbf{e}_{bf}^{(\ell)}$ and $\mathbf{e}_{uf}^{(\ell)}$ are similar to $\mathbf{e}_{b}$ and $\mathbf{e}_{u}$ in Equation \ref{eq:f2}. Then we do as Equation \ref{eq:f3} and \ref{eq:f4} to get a rate $\mathbf{r}$. We select the bundle $\mathit{b_t}$ with largest $\mathbf{r}$ as base. For every other bundle $\mathit{b}$ we compute a factor $\eta_{b} = \left|\mathcal{I}_{b} \cap \mathcal{I}_{b_t}\right| /\left|\mathcal{I}_{b_t}\right| $, where $\mathcal{I}_{b}$ is the item set for bundle $b$. 

\subsection{Information Aggregation and Prediction}\label{sec:infoap}
\noindent After obtaining $\mathbf{e}_{u, 1}^{(\ell+1)}$, $\mathbf{e}_{u, 2}^{(\ell+1)}$, $\mathbf{e}_{b, 1}^{(\ell+1)}$ and $\mathbf{e}_{b, 2}^{(\ell+1)}$, we firstly average them to get $\mathbf{e}_{u}^{(\ell+1)}$ and $\mathbf{e}_{b}^{(\ell+1)}$. In real implementation, we have in total $L=4$ layers of information propagation and could get five stages of embeddings. To utilize the whole information we use aggregation function to synthesize all the intermediate embeddings. Here we adopt concatenation as aggregation function:
\begin{small}
\begin{equation}\label{eq:f2}
\begin{aligned}
&\mathbf{e}_{u} =  [\mathbf{e}_{u}^{0} \thinspace ||\thinspace \mathbf{e}_{u}^{1} \thinspace||\thinspace \mathbf{e}_{u}^{2}\thinspace||\thinspace \mathbf{e}_{u}^{3}\thinspace ||\thinspace \mathbf{e}_{u}^{4}]\\
&\mathbf{e}_{b} =  [\mathbf{e}_{b}^{0}\thinspace ||\thinspace \mathbf{e}_{b}^{1}\thinspace ||\thinspace \mathbf{e}_{b}^{2}\thinspace||\thinspace \mathbf{e}_{b}^{3} \thinspace||\thinspace \mathbf{e}_{b}^{4}]
\end{aligned}
\end{equation}
\end{small}

\begin{small}
\begin{equation}\label{eq:f3}
\mathbf{e}_{sub} =  [\mathbf{e}_{u} \thinspace || \thinspace \mathbf{e}_{b}]
\end{equation}
\end{small}

$\mathbf{e}_{sub}$ is the graph-level embedding indicating the correlation between target user \textit{u} and target bundle \textit{b}. This embedding contains information about the user's preference on this bundle. Then as shown in Figure \ref{tab:overall_graph}, we use two linear layers to get the final prediction between 0 and 1. For the first layer, we choose \textit{Relu} as the activation function and then use a \textit{Sigmoid} after the second linear layer to ensure the preference score is between 0 and 1. The procedure is as follows:
\begin{equation} \label{eq:f4}
\mathbf{r}= \textrm{Sigmoid}(\mathbf{W}_{mlp2} \cdot \textrm{Relu}(\mathbf{W}_{mlp1} \mathbf{e}_{sub}))
\end{equation}

\subsection{Label Leakage Issue}
\noindent As proposed by \cite{jiani2019leakage} and \cite{deng2020personalized}, label leakage is a common but implicit issue on GNN models. Label leakage issue happens when predicting links by aggregating information from nearby nodes and the target link is also included in the aggregating function. In this case, the mapping learned by model will have a self-mapping problem. Specifically, if we try to predict an edge \textit{q} between target bundle \textit{b} and target user \textit{u}, then in every iteration bundle \textit{b} aggregates information from \textit{u} and user \textit{u} also aggregates information from \textit{b}. Therefore, the model tends to learn a self mapping $f_{\theta} (\textit{q},...) = \textit{q}$ \cite{deng2020personalized}. This is not an ideal mapping function, since the information of target bundle \textit{b} and target user \textit{u} are lost as their embeddings are smoothed by each other, and the model tends to become overfitting.
\par In our \suger\ model, this problem is even more critical since the core of \suger\ is to use user-bundle centered subgraph to predict the center edge. Although we apply different weight matrices for user $\rightarrow$ bundle and bundle $\rightarrow$ user, this could not solve the problem thoroughly.
\par The key solution to this label leakage issue is to ensure the edge we predict is not involved in the information propagation process. To achieve this, we delete edges between centered user and bundle when training the model to cut off such self-mapping loops and to ensure the information is not propagated between centered bundle \textit{b} and centered user \textit{u} when the model learns how to predict the edge $\textit{q} = (\textit{b},\textit{u}) $ itself. Compared to the dropout layer solution, it is safer than randomly dropping some edges since deletion can ensure the label leakage issue does not happen between target user and bundle. Another disadvantage of randomly dropping is that it might negatively change graph structure. For example, as Table \ref{tab:datasetsta} suggests, the datasets of Netease and Youshu are sparse. The generated subgraphs are usually not dense, so such a dropout layer could easily change the graph structure. 

\subsection{Transfer Bundle Recommendation}
\noindent One important contribution of this paper is that our model outperforms other bundle recommendation models in the transfer learning context as shown in table \ref{tab:transfer_result}. We train the model on one dataset and then use testset from another dataset which is unseen by the model during training stage. The performance is still acceptable compared to the previous model like BGCN \cite{chang2021rec}. As mentioned above, \suger will generate k-hop subgraph to learn the local graph pattern of a pair of bundle and user. There is no side information like user's personal info is used during the initialization stage. We hope this model could learn something about user's internal purchase logic which is independent of the platform. For example, if you are more likely to buy a bundle with only one familiar element then this pattern will appear in all online shopping platforms you visit. In the graph context, all such locally centered subgraphs will have one edge with high probability and other edges with relatively low probability during predicion.

\section{Experiments}\label{sec:experiments}

\noindent In this section, we conduct experiments to verify the effectiveness and transferability of \suger\ model.

\subsection{Experimental Setting and Dataset Statistics}

\noindent \textbf{Dataset Statistics.} The datasets we use in this paper are Netease and Youshu, and the statistics can be found in Table \ref{tab:datasetsta}. Netease dataset contains the music bundle data and purchase record from Netease Inc and is collected by \cite{netease}. Youshu is a famous chinese book review website, and the dataset contains book bundles and user ratings. The dataset is collected by \cite{liang2019rec}. From Table \ref{tab:datasetsta} we can see Netease is larger and sparser than Youshu.

\begin{table}[tb]\small
\centering
\caption{Statistics of two datasets}
\begin{tabular}{|l|l|l|l|l|l|}
\hline
\textit{Dataset} & User  & Bundle & Item   & U-B    & U-I     \\ \hline
\textit{Youshu}  & 8039  & 4771   & 32770  & 51377  & 138515  \\ \hline
\textit{Netease} & 18528 & 22684  & 123628 & 302303 & 1128065 \\ \hline
\end{tabular}
	\label{tab:datasetsta}
\end{table}



			

			



\noindent \textbf{Baselines.} We use four baselines for the overall performance comparison experiment, they are \textbf{GCN-BG} \cite{gcn}, \textbf{NGCF-BG} \cite{ngcf}, \textbf{DAM} \cite{liang2019rec} and \textbf{BGCN} \cite{chang2021rec}. For \textbf{GCN} \cite{gcn} and \textbf{NGCF} \cite{ngcf}, both of them have two kinds (Bipartite-Graph and Tripartite-Graph ) of implementations and we choose BG in our experiments by using user-bundle bipartite graph to train and test the performance.
For these four baselines, \textbf{GCN-BG} \cite{gcn} and \textbf{NGCF-BG} \cite{ngcf} are GNN models for traditional recommendation tasks, and are not designed specifically for bundle recommendation. \textbf{DAM} \cite{liang2019rec} and \textbf{BGCN} \cite{chang2021rec} are newly published bundle-oriented GNN models for recommendation. 

\noindent \textbf{Metric.} We use two widely used metrics \textit{Recall@K} \cite{ngcf} and \textit{NDCG@K} \cite{ndcg} to test the performance of all experiments. \textit{Recall@K} measures the ratio of real positive bundles in top-$K$ bundles. \textit{NDCG@K} takes the rank position into consideration and gives higher scores to the real positive bundle at a higher rank in the top-$K$ list.

\noindent \textbf{Experimental Setting.} For all experiments, we first split Youshu and Netease datasets into training and test sets. We split the positive user-bundle interactions into two parts for training and test, so during training, the model cannot use the positive u-b pair from the test split. For both of them, roughly about 60\% of the user-bundle interactions are in the training set. We use random negative sampling for training, and the positive-negative ratio is one, so the number of triplets is the same as the number of positive samples in the training set. 
\par During training stage, we use Adam optimizer with learning rate $3 \mathrm{e}-5$ and weight decay $2 \mathrm{e}-7$. For the regularization factor in our BPR loss function, we search in the range $\{1 \mathrm{e}-1,1 \mathrm{e}-2,1 \mathrm{e}-3,1 \mathrm{e}-4,1 \mathrm{e}-5,1 \mathrm{e}-6,1 \mathrm{e}-7 \}$ and choose $1\mathrm{e}-5$.

\begin{table*}[tb]\small
	\centering
	\caption{Performance compared with four baselines on two widely-used datasets}
 \vspace{-1.0\baselineskip}	\renewcommand\arraystretch{1.5} 
	\begin{threeparttable}
        \resizebox{\linewidth}{!}{
		\begin{tabular}{c|c|c|c|c|c|c|c|c|c|c|c|c}
			\toprule[0.3pt]
            \multirow{2}{*}{\textbf{Model}} &  \multicolumn{6}{c|}{\textbf{Youshu}} & \multicolumn{6}{c}{\textbf{Netease}}\\
			
			\cline{2-7} \cline{8-13} 
			& \textbf{Recall@20} & \textbf{Recall@40} & \textbf{Recall@80}  & \textbf{NDCG@20} & \textbf{NDCG@40} & \textbf{NDCG@80} & \textbf{Recall@20} & \textbf{Recall@40} & \textbf{Recall@80}  & \textbf{NDCG@20} & \textbf{NDCG@40} & \textbf{NDCG@80} \\
			
			\midrule 
			\midrule
			
			\textit{NGCF-BG} & 0.1343 & 0.2076 & 0.2447 & 0.0618 & 0.0925 & 0.1420 & 0.0551 & 0.0927 & 0.1356 & 0.0265 & 0.0464 & 0.0768\\
			\hline

			\textit{GCN-BG} & 0.1431 &	0.2191 & 0.2432&	0.0739&	0.1137&	0.1652	&0.0567&	0.0982&	0.1439&	0.0292&	0.0501&	0.0799\\
			\hline

			\textit{DAM} & 0.1670&	0.2086&	0.2918&	0.0757&	0.0939&	0.1532	&	0.0747&	0.1290&	0.1695&	0.0351&	0.0593&	0.0840\\
			\hline

			\textit{BGCN} & \underline{0.2531}&	\underline{0.3674}&	\underline{0.4755}&	\underline{0.1623}&	\underline{0.1858}&	\underline{0.2432}	&	\underline{0.1494}&	\underline{0.1825}&	\underline{0.2421}&	\underline{0.0786}&	\underline{0.1069}&	\underline{0.1428}
			\\
			\hline

			\textit{\suger} & \textbf{0.3529}&	\textbf{0.5438}&	\textbf{0.6682}&	\textbf{0.2041}&	\textbf{0.2961}&	\textbf{0.3960}	&	\textbf{0.2835}&	\textbf{0.3539}&	\textbf{0.4301}&	\textbf{0.1361}&	\textbf{0.1894}&	\textbf{0.2359}
			\\
			\hline

			\hline
            \textit{\%Improv.} & 39.43\% & 	48.01\% & 	40.53\% & 	25.75 \% &	59.36 \% &	62.83 \% &		89.76\% & 	93.92 \% &	77.65\% & 	73.16\% & 	77.17\% & 	65.20 \% 
			\\
			\hline

			\bottomrule[1.2pt]
		\end{tabular}}
	\end{threeparttable}
	\label{tab:all_result}
\end{table*}

\subsection{Overall Performance Comparison}

\noindent We do experiments on Youshu and Netease datasets with four baselines as listed above. We use \textit{Recall@K} and \textit{NDCG@K} as metrics, and choose three different \textit{K} among 20,40 and 80 which are the same as \cite{chang2021rec}. The overall performance is recorded in Table \ref{tab:all_result}.
\par For the result, we have a brief analysis. First, \suger\ model outperforms all the baselines with a large improvement when using \textit{Recall@K} and \textit{NDCG@K} as the metrics. We can see that \suger\ could get an average of 45\% improvement on Youshu and an average of 79\% improvement on Netease compared to the state-of-the-art model \textbf{BGCN}\cite{chang2021rec}. That proves the effectiveness of our user-bundle centered subgraph-based GNN model.

\begin{table*}[tb]\small
	\centering
	\caption{Transfer Bundle Recommendation Experiment}
 \vspace{-1.0\baselineskip}	\renewcommand\arraystretch{1.5} 
	\begin{threeparttable}
        \resizebox{\linewidth}{!}{
		\begin{tabular}{c|c|c|c|c|c|c|c|c|c|c|c|c}
			\toprule[0.3pt]
            \multirow{2}{*}{\textbf{Model}} &  \multicolumn{6}{c|}{\textbf{Youshu}} & \multicolumn{6}{c}{\textbf{Netease}}\\
			
			\cline{2-7} \cline{8-13} 
			& \textbf{Recall@20} & \textbf{Recall@40} & \textbf{Recall@80}  & \textbf{NDCG@20} & \textbf{NDCG@40} & \textbf{NDCG@80} & \textbf{Recall@20} & \textbf{Recall@40} & \textbf{Recall@80}  & \textbf{NDCG@20} & \textbf{NDCG@40} & \textbf{NDCG@80} \\
			
			\midrule 
			\midrule

			\textit{GCN-BG} & 0.0611&	0.0925&	0.1233&	0.0305&	0.0474&	0.0735&		0.0244&	0.0415&	0.0639&	0.0121&	0.0207&	0.0348

			\\
			\hline

			\textit{BGCN} & 0.1801&	0.2809&	0.3476&	0.1354&	0.1419&	\textbf{0.1876}&		\textbf{0.1061}&	0.1496&	0.1819&	\textbf{0.0559}&	0.0848&	0.1014

			\\
			\hline

			\textit{\suger} & \textbf{0.3107}&	\textbf{0.3926}&	\textbf{0.4255}&	\textbf{0.1537}&	\textbf{0.1725}&	0.1801	&	0.0955&	\textbf{0.1706}&	\textbf{0.2302}&	0.0408&	\textbf{0.0850}&	\textbf{0.1374}
			\\
			\hline

			\bottomrule[1.2pt]
		\end{tabular}}
	\end{threeparttable}

	\label{tab:transfer_result}
\end{table*}

\subsection{Transfer Bundle Recommendation Experiment}

\noindent We first train two models on the training sets of Youshu and Netease, then conducting test on the test sets of Youshu using Netease's model and test on the test set of Netease using Youshu's model. We adopt the same setting on \textbf{BGCN}\cite{chang2021rec} and \textbf{NGCF}\cite{ngcf}. The result is shown in Table \ref{tab:transfer_result}. We can observe that \suger\ has the best transfer performance compared to two baselines. Another observation is that the performance of the transfer setting still decreases compared with the performance in Table \ref{tab:all_result}. This experimental results follows our intuition that the subgraph-based bundle recommendation model we propose  has strong transferability, which is elaborated in Section \ref{sec:initialization}. The result shows that the model is able to learn the internal purchasing logic, which is not platform-dependent. As a result, the heterogeneous subgraph we generate could contain some platform-independent semantics. 

\begin{table*}[tb]\small
	\centering
	\caption{Ablation Experiment}
 \vspace{-1.0\baselineskip}	\renewcommand\arraystretch{1.5} 
	\begin{threeparttable}
        \resizebox{\linewidth}{!}{
		\begin{tabular}{c|c|c|c|c|c|c|c|c|c|c|c|c}
			\toprule[0.3pt]
            \multirow{2}{*}{\textbf{Variant}} &  \multicolumn{6}{c|}{\textbf{Youshu}} & \multicolumn{6}{c}{\textbf{Netease}}\\
			
			\cline{2-7} \cline{8-13} 
			& \textbf{Recall@20} & \textbf{Recall@40} & \textbf{Recall@80}  & \textbf{NDCG@20} & \textbf{NDCG@40} & \textbf{NDCG@80} & \textbf{Recall@20} & \textbf{Recall@40} & \textbf{Recall@80}  & \textbf{NDCG@20} & \textbf{NDCG@40} & \textbf{NDCG@80} \\
			
			\midrule 
			\midrule

			\textit{Base-transfer} & \textbf{0.3107}&	\textbf{0.3926}&	\textbf{0.4255}&	\textbf{0.1537}&	\textbf{0.1725}&	\textbf{0.1801}&		\textbf{0.0955}&	\textbf{0.1706}&	\textbf{0.2302}&	
			\textbf{0.0408}&	\textbf{0.0850}&	\textbf{0.1374}

			\\
			\hline

			\textit{No subgraph-transfer} & 0.0303&	0.0348&	0.0472&	0.0149&	0.0170&	0.0204	&	0.0271	&0.0369	&0.0457&	0.0112&	0.0166&	0.0217

			\\
			\hline

			\textit{Leakage-transfer} & 0.0714&	0.1049&	0.1725&	0.0301&	0.0522&	0.0913	&	0.0593&	0.1104&	0.1349&	0.0312&	0.0530&	0.0645

			\\
			\hline

			\textit{No subgraph} & 0.0301	 & 0.0353 & 	0.0469 & 	0.0141	 & 0.0176 & 	0.0218 & 		0.0315 & 	0.0349 & 	0.0392 & 	0.0157 & 	0.0173 & 	0.0195
			\\
			\hline

			\textit{Leakage} & 0.1301&	0.2041&	0.2886&	0.0637&	0.0912&	0.1475	&	0.0692&	0.1345&	0.1688&	0.0349&	0.0621&	0.0837
			\\
			\hline

			\bottomrule[1.2pt]
		\end{tabular}}
	\end{threeparttable}
		\label{tab:ablation_result}
\end{table*} 

\vspace{-1.0\baselineskip}
\begin{figure}[htbp]
    \small
    \centering
    \includegraphics[width=9cm,height=3.7cm]{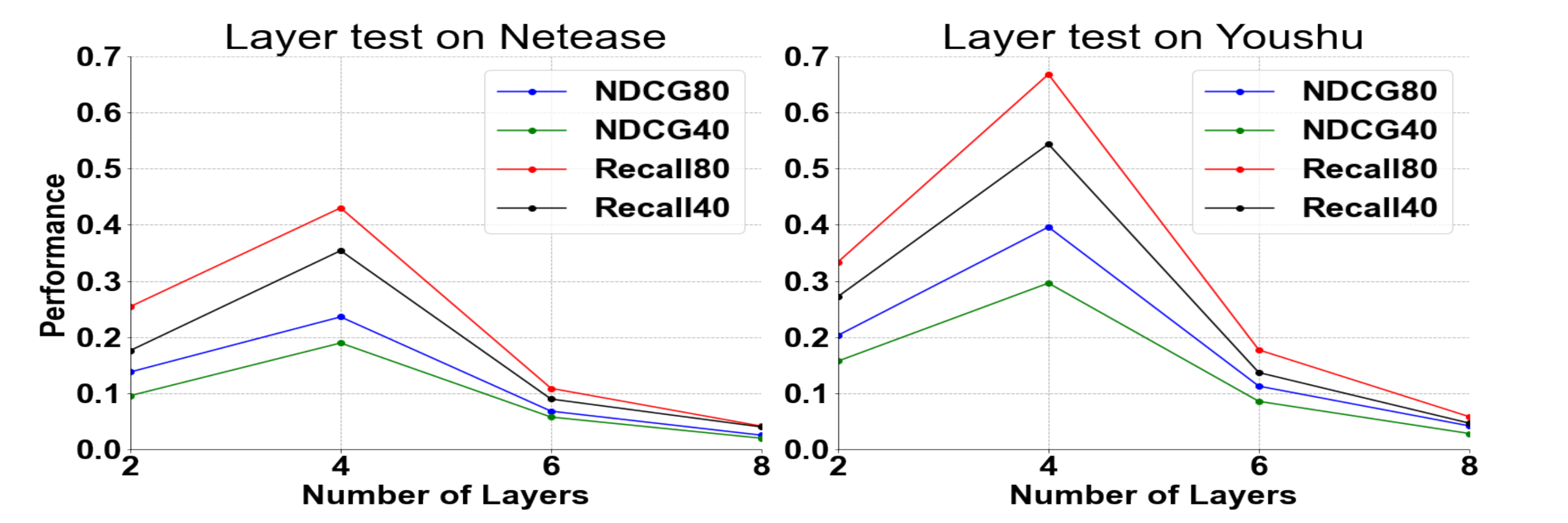} 
    \vspace{-1.0\baselineskip}
    \caption{Result of Layer test on two datasets}\label{tab:layer_graph}
\end{figure}

\subsection{Ablation and Hyperparameter Study}
\noindent We do ablation experiments on the label leakage issue, number of convolution layers, usage of subgraph and effectiveness of transfer model. The result is shown in Table \ref{tab:ablation_result}, and Figure \ref{tab:layer_graph}. 
\noindent \textit{Base-transfer} means we train a model using transfer setting as mentioned in the previous section . \textit{No subgraph} means we do not generate $k$-hop heterogeneous user-bundle subgraph as input but only initialize embeddings for training and test sets. \textit{Leakage} means we do not delete target edges when training. \textit{No subgraph-transfer} and \textit{Leakage-transfer} are the same as \textit{No subgraph} and \textit{Leakage} while using transfer setting for training and test. 
\noindent We can see the label leakage issue will largely affect the performance since the model will have self-mapping issue as illustrated in Section \ref{sec:infoap}. Another observation is that subgraph input is important since the model cannot capture structure information without it, resulting in low performance. 
\noindent We further conduct a experiment to evaluate the impact of the number of layers for GNN aggregation. The result is shown in Figure \ref{tab:layer_graph}. We could see that 4-layer model has the best result, and the performance starts to drop with larger number of layers.

\section{Related work}\label{sec:related-work}

\noindent \textbf{Bundle Recommendation.}
Bundle recommendation is different from traditional user-item recommendation, so the classic model like "Two-Tower model" \cite{yi2019rec} could not be applied directly to solve the problem. The research about bundle recommendation is not thorough and still in the initial stage even the concept of "bundle" is becoming more popular among e-commerce and game companies. Some GNN based traditional recommendation models could also be applied in bundle context such as GCN \cite{gcn}, and NGCF \cite{ngcf}. These models focus on extracting information from the homogeneous bipartite graph but it cannot achieve high performance on bundle context. Another type of model is bundle-oriented, like DAM \cite{liang2019rec} and BGCN \cite{chang2021rec} where the latter is the state-of-the-art model.


\section{Conclusion}\label{sec:conclusion}
\noindent In this paper, we propose a subgraph-based GNN model to handle both the basic bundle recommendation and transfer bundle recommendation problem. The model uses three interaction matrices as input to generate \textit{k}-hop heterogeneous user-bundle centered subgraph to learn the embeddings for target bundles and users. The model has strong transferability when facing unseen domains. We also propose a solution for the implicit label leakage issue in our model to avoid the self-mapping problem. Extensive experiments demonstrate the significant improvement of the effectiveness and transferability of our model over all the baseline models. 


\bibliographystyle{style_files/named}
\bibliography{main.bib}


\end{document}